\documentclass[preprint,superscriptaddress,showkeys,showpacs,nofootinbib,prd,byrevtex]{revtex4} 
\usepackage{graphicx}
\usepackage{bm}
\begin{document}      
\preprint{CYCU-HEP-09-06}
\preprint{YITP-08-13}
\preprint{INHA-NTG-07/2008}
\title{Magnetic susceptibility of the QCD vacuum\\
 at finite quark-chemical potential}      
\author{Seung-il Nam}
\email[E-mail: ]{sinam@cycu.edu.tw, sinam@yukawa.kyoto-u.ac.jp}
\affiliation{Department of Physics, Chung-Yuan Christian University, Chung-Li 32023, Taiwan} 
\affiliation{Yukawa Institute for Theoretical Physics, Kyoto University, 
\\Kyoto 606-8502, Japan} 
\author{Hui-Young Ryu}
\email[E-mail: ]{hyryu@rcnp.osaka-u.ac.jp} 
\affiliation{Research Center for Nuclear Physics, Ibaraki,
Osaka 567-0047, Japan} 
\author{M.~M.~Musakhanov}
\email[E-mail: ]{yousuf@uzsci.net}
\affiliation{Theoretical Physics Department, Uzbekistan
University, Tashkent 7000174, Uzbekistan} 
\author{Hyun-Chul Kim}
\email[E-mail: ]{hchkim@inha.ac.kr}
\affiliation{Department of Physics, Inha University,
Incheon 402-751, Korea} 
\date{April 2009}
\begin{abstract}
We investigate the QCD magnetic susceptibility $\chi$ at the finite
quark-chemical potential ($\mu\neq 0$) and at zero temperature ($T=0$)
to explore the pattern of the magnetic phase transition of the QCD 
vacuum. For this purpose, we employ the nonlocal chiral quark model
derived from the instanton vacuum in the presence of the chemical
potential in the chiral limit.  Focusing on the Nambu-Goldstone phase,
we find that the magnetic susceptibility remains almost stable to
$\mu\approx200$ MeV, and falls down drastically until the the
quark-chemical potential reaches the critical point $\mu_c\approx320$
MeV.  Then, the strength of the $\chi$ is reduced to be about a half
of that at $\mu=0$, and the first-order magnetic phase transition
takes place, corresponding to the chiral restoration.  From these
observations, we conclude that the response of the QCD vacuum becomes
weak and unstable to the external electromagnetic field near the
critical point, in comparison to that for vacuum.  It is also shown
that the breakdown of Lorentz invariance for the magnetic
susceptibility, caused by the finite chemical potential, turns out to 
be small.  
\end{abstract}
\maketitle
\section{Introduction}
It is of great importance to understand nontrivial structures of the
QCD vacuum,  since it reflects strong nonperturbative fluctuations and
lay the foundation for explaining low-energy phenomena of hadrons. 
In particular, the quark condensate characterises the nonperturbative 
features of the QCD vacuum and plays a role of an order parameter
associated with spontaneous breakdown of chiral symmetry (SB$\chi$S)
which is known to be essential in low-energy hadronic
phenomena.  The nature of the QCD vacuum can be examined under
external fields such as electromagnetic (EM) field, since it
reveals the EM properties of the vacuum.  The magnetic
susceptibility is defined as the vacuum expectation value (VEVs) in
Euclidean space:  
\begin{equation}
\label{eq:VEV}
\langle{iq}^{\dagger}\sigma_{\mu\nu}q\rangle_{F}
=ie_q{F_{\mu\nu}}\langle{iq}^{\dagger}q\rangle \chi,
\end{equation}
where $e_q$ denotes the electric charge of the quark and $F_{\mu\nu}$
represents the EM field-strength tensor.  The $\langle i q^{\dagger} q
\rangle$ stands for the quark (chiral) condensate taken as a
normalization. It is natural to take it as a normalization for the
magnetic susceptibility, since the VEV
$\langle{iq}^{\dagger}\sigma_{\mu\nu}q\rangle_{F}$ breaks chiral 
symmetry~\cite{Ioffe:2005ym}.  Recently, Braun et
al.~\cite{Braun:2002en} have proposed that the $\chi$ may be measured
in the exclusive photoproduction of hard dijets $\gamma+N\to 
q\bar{q}+N$. Theoretically, there have been already many 
works on it in the QCD sum rules and chiral effective
models~\cite{Belyaev:1984ic,Balitsky:1985aq,Ball:2002ps, 
Kim:2004hd,Dorokhov:2005pg,Rohrwild:2007yt,Goeke:2007nc}.  The
magnetic susceptibility is also related to the photon distribution 
amplitude~\cite{Ball:2002ps}.  
    
While the magnetic susceptibility is relatively well understood in
free space, it is not much known about its modification at finite
density (finite quark-chemical potential, $\mu$).  For example, there
are several interesting theoretical points as follows: First, in
principle, the pattern of the {\it magnetic} phase transition of QCD
vacuum can be revealed by studying the change of the magnetic
susceptibility in medium~\cite{Sato:2008js}.  Second, one can see how
the vacuum reacts in the presence of the external EM source,
especially, near the critical point.  Third, the medium modification
of the $\chi$ may play a critical role in explaining the exclusive
photoproduction of the hard dijets in nuclei.  Finally, this
information may be important in understanding 
the EM features of neutron stars~\cite{Tatsumi:2008nx}.  Moreover,
since Lorentz invariance is broken in medium, we must consider
separately the space and time components of the magnetic
susceptibility at finite quark-chemical potential.  In fact,
Refs.~\cite{Nam:2008xx,Kim:2003tp,Kirchbach:1997rk} have studied the
importance of the breakdown of Lorentz invariance for the in-medium
pion weak decay constant $F_\pi$.  In fact, Ref.~\cite{Nam:2008}
has calculated the magnetic susceptibility without the breakdown of
Lorentz invariance taken into account.  In addition, the chemical
potential was treated perturbatively in Ref.~\cite{Nam:2008}.  We will
consider explicitly the breakdown of Lorentz invariance in the present
work.   
 
In order to compute the $\chi$ at finite quark-chemical potential, we employ the 
framework of the $\mu$-modified nonlocal chiral quark model
(NL$\chi$QM) derived from the instanton vacuum~\cite{Carter:1998ji}.   
The NL$\chi$QM has several virtues: All relevant QCD symmetries are 
satisfied within the model. In particular, the instanton vacuum well
realizes spontaneous breakdown of chiral symmetry via quark zero
modes. Moreover, there are only two parameters: the average size of
instantons ($\bar{\rho}\approx 1/3$ fm) and average inter-instanton
distance ($\bar{R}\approx1$ fm), which can be determined by the
internal constraint such as the self-consistent
equation~\cite{Shuryak:1981ff,Diakonov:1983hh,Diakonov:1985eg}. These   
values for $\rho$ and $R$ have been supported in various LQCD
simulations recently~\cite{Chu:vi,Negele:1998ev,DeGrand:2001tm}.   
There is no further adjustable parameter in the model.  Note that the
inverse of the average size of instantons, $\bar{\rho}^{-1}\simeq
\,600$ MeV, is regarded as the normalization point of the present
model, so that we can calculate the dependence of the magnetic
susceptibility on the chemical potential to the critial point
($\mu\approx 320$ MeV) without any trouble.  

We concentrate in this work on the Nambu-Goldstone (NG) phase.  In 
order to go beyond the NG phase, we need subtle techniques which we
will explain briefly later.  As a result, we find that the $\chi$
remains stable up to $\mu\approx200$ MeV, and falls down drastically
to the critical quark-chemical potential, $\mu_c\approx320$ MeV and
the strength of the $\chi$ is reduced to be about a half of that at
$\mu=0$.  The {\it first-order} magnetic phase transition takes 
place at this critical point corresponding to the phase of the chiral
restoration.  We will show that the response of the QCD vacuum becomes
weak and unstable to the external EM source near the critical point.
Moreover, the breakdown of Lorentz invariance turns out to be small in
contrast to that in the case of the pion weak decay constant 
$F_\pi$~\cite{Nam:2008xx}.  

We organize the present work as follows: In Section II, we briefly
explain the $\mu$-modified NL$\chi$QM.  In Section III, we compute the
magnetic susceptibility $\chi$ and present the numerical results with 
discussions.  The final Section summarises the present work and draws 
conclusions.  The outlook of this work is also shortly discussed in
this Section.

\section{Nonlocal chiral quark model at finite quark-chemical potential}
A modified Dirac equation with the finite quark-chemical
potential $\mu$ in (anti)instanton ensemble can be written in
Euclidean space as follows:   
\begin{equation}
\label{eq:MDE}
\left[i\rlap{/}{\partial}-i\rlap{/}{\hat{\mu}}+\rlap{/}{A}_{I\bar{I}}
\right]\Psi^{(n)}_{I\bar{I}}  
=\lambda_n\Psi^{(n)}_{I\bar{I}},
\end{equation}
where $\hat{\mu}_\alpha=(0,0,0,\mu)$. Note that we consider here the chiral
limit ($m\to0$). The subscript $(\bar{I})I$ stands for the
(anti)instanton contribution, and we use a singular-gauge instanton 
solution:   
\begin{eqnarray}
\label{eq:instanton}
A^{\alpha}_{\mu}(x)
=\frac{2\bar{\eta}^{\alpha\nu}_{\mu}\bar{\rho}^2x_{\nu}}
{x^2(x^2+\bar{\rho}^2)},
\end{eqnarray}
where $\eta^{\alpha\nu}_{\mu}$ and $\bar{\rho}$ denote the 'tHooft
symbol and average instanton size ($\bar{\rho}\approx1/3$ fm),
respectively. Since the instanton has its own spatial distribution
with finite size ($\rho$) in principle, one needs to consider an 
instanton-distribution function when the integral over the 
collective coordinate for the instanton (instanton center, color
orientation, and instanton size) is
performed~\cite{Shuryak:1981ff,Diakonov:1983hh,Diakonov:1985eg}. However,  
we assume here a simple $\delta$-fucntion type for the
insatanton-distribution function, i.e., $\delta(\rho-\bar{\rho})$.  In
fact, the effect of the finite size is taken as $1/N_c$
corrections~\cite{Diakonov:1983hh,Diakonov:1985eg}.  We refer to
Ref.~\cite{Goeke:2007nc} for the finite-size effect on the magnetic
susceptibility in free space.  The quark zero-mode solution 
then can be obtained by solving the following equation: 
\begin{equation}
\label{eq:ZM}
\left[i\rlap{/}{\partial}-i\rlap{/}{\hat{\mu}}
+\rlap{/}{A}_{I\bar{I}}\right]\Psi^{(0)}_{I\bar{I}}
=0.
\end{equation}
The explicit form of $\Psi^{(0)}$ can be found in
Refs.~\cite{Abrikosov:1981qb,Carter:1998ji}. Assuming that the
low-energy hadronic properties are dominated by the zero mode, we can
write the quark propagator in one instanton background approximately
as follows:   
\begin{eqnarray}
S_{I\bar{I}}(x,y) & = & \langle \psi(x)\psi(y)^\dagger\rangle = -\sum_n 
\frac{\Psi_{I\bar{I}}^{(n)}(x)\Psi_{I\bar{I}}^{(n)\dagger}(y)}
{\lambda_n+im}\approx S_0(x,y) -
\frac{\Psi^{(0)}_{I\bar{I}}(x)
  \Psi^{(0)\dagger}_{I\bar{I}}(y)}{im},    
\label{eq:prop1}
\end{eqnarray}
where $S_0$ is a free quark propagator, $S_0=(i\rlap{/}{\partial} -
i\rlap{/}{\hat{\mu}})^{-1}$. Incorporating Eqs.~(\ref{eq:instanton}) and
(\ref{eq:prop1}), and averaging over all instanton collective
coordinates, we can obtain the following expression for the quark
propagator in momentum space:    
\begin{equation}
\label{eq:prop2}
S(p,\mu)=-\frac{1}{\rlap{/}{p}+i\rlap{/}{\hat{\mu}}-iM(p,\hat{\mu})}. 
\end{equation}
The momentum- and $\mu$-dependent effective quark mass $M(p,\hat{\mu})$,
corresponding to the Fourier transform of the quark zero-mode solution 
of Eq.~(\ref{eq:ZM}), reads: 
\begin{equation}
\label{eq:MDQM}
M(p,\hat{\mu})=M_0(\hat{\mu})\,[(p+i\hat{\mu})^2\psi^2(p,\hat{\mu})], 
\end{equation}
where $M_0$ designates the constituent-quark mass as a function of
$\hat{\mu}$, which will be determined self-consistently within the
model. The analytical expression for $\psi(p,\hat{\mu})$ is also given
in Ref.~\cite{Carter:1998ji}. In actual calculations, we make use of a
dipole-type parameterization, instead of using Eq.~(\ref{eq:MDQM}),
for numerical simplicity: 
\begin{eqnarray}
\label{eq:MFD}
M(p,\hat{\mu})&\approx&M_0(\hat{\mu})
\left[\frac{2\Lambda^2}{(p+i\hat{\mu})^2+2\Lambda^2}
\right]^2.
\end{eqnarray}
The cutoff mass $\Lambda$ corresponds to the
inverse of the average (anti)instanton size ($\sim1/\bar{\rho}$),
resulting in $\Lambda\approx600$ MeV. We have verified that this
simple parameterization reproduces Eq.~(\ref{eq:MDQM}) qualitatively
well~\cite{Nam:2008fe}.  One can refer to Ref.~\cite{GomezDumm:2006vz}  
for more quantitative studies on the parameterization of
$M(p,\hat{\mu})$.   

Now we are in a position to consider the effective partition function
for $N_f=2$ from the instanton vacuum configuration with the finite
$\mu$ in the large $N_c$ limit. Choosing the relevant terms for
further discussion, it can be written as follows~\cite{Carter:1998ji}:  
\begin{eqnarray}
\label{eq:PF}
\mathcal{Z}_{\rm eff} & = & \int{d\lambda}\,{D\psi}\,{D\psi^{\dagger}}
\exp\left[\int d^4x\,\psi^{\dagger}(i\rlap{/}{\partial}
  -i\rlap{/}{\hat{\mu}}+im)\psi+\lambda(Y^++Y^-)-N\ln\lambda\right],   
\end{eqnarray}
where the flavour, colour, and Dirac spin indices are assumed tacitly.
Note that this effective partition function has been constructed in
such way that we can obtain the quark propagator given in 
Eq.~(\ref{eq:prop1}). $Y^{\pm}$ denotes the $2N_f$-'tHooft interaction  
in the (anti)instanton background at finite $\mu$. As for the
parameters appearing in the second and third terms of
Eq.~(\ref{eq:PF}), $\lambda$ plays a role of the Lagrangian
multiplier, whereas $N$ indicates the sum of the number of 
the (anti)instantons.  For more details on this partition function and
its derivation, one may refer to Ref.~\cite{Carter:1998ji}, and
references therein.  

We would like to explain briefly the phase structure obtained via the
present model. Since both the quark-antiquark and quark-quark
interactions for $N_f=2$ are attractive, two possible phases arise: 
The Nambu-Goldstone (NG) and colour-superconducting (CSC) phases.  They 
are characterised by the QCD order paramters, i.e., the chiral and 
diquark condensates, respectively, or by the finite 
constituent quark mass $M_0$ and diquark energy gap $\Delta$,
equivalently.  As shown in Ref.~\cite{Carter:1998ji}, solving the
Dyson-Schwinger-Gorkov (DSG) equation and using the instanton packing
fraction $N/V\approx(200\,\mathrm{MeV})^4$, one can compute $M_0$ and
$\Delta$ as functions of $\mu$ by minimizing the partition function
with respect to $\lambda$. In the present work, we consider only the
pure NG phase, in which the magnetic susceptibility is well defined
in terms of the chiral condensate as shown in Eq.~(\ref{eq:VEV}). We
ignore, for simplicity, the metastable mixed phases.  As a
result, the value of the critical chemical potential ($\mu_c$) was
determined to be about $320$ MeV. We note that it is about 
$5\%$ different from that given in Ref.~\cite{Carter:1998ji} in which
the renormalization scale is taken to be slightly different from ours,  
$\Lambda\approx1/\bar{\rho}\approx0.6$ GeV.

\section{QCD magnetic susceptibility at finite $\mu$}
In this Section, we compute the VEV of the tensor current of
Eq.~(\ref{eq:VEV}) in the the background EM field in the presence of
the finite $\mu$ and provide numerical results for the magnetic
susceptibility $\chi$.  In order to deal with the background EM field,
we employ the Schwinger
method~\cite{Kim:2004hd,Schwinger:1951nm}. Then, we can write the
covariant effective partition function with an external tensor source
$T$ as follows:     
\begin{eqnarray}
\label{eq:ECA}
\mathcal{Z}_{\mathrm{eff}} [T,\mu] &=&
\int{d\lambda}\,{D\psi}\,{D\psi^{\dagger}} 
\exp\left[\sum^2_{f=1}\int
  d^4x\,\psi^{\dagger}_f(i\rlap{/}{D}-i\rlap{/}{\hat{\mu}}+\sigma\cdot 
  T)\psi_f +\lambda(Y^+_2+Y^-_2)\right]. 
\end{eqnarray}
where we have written only the relevant terms to compute the
VEV. $iD_\mu$ indicates the covariant derivative, $i\partial_\mu
+ e_qV_\mu$, in which $e_q$ and $V_\mu$ are the quark electric charge  
and external photon field, respectively.  The $\sigma_{\mu\nu}$ is the 
well-known spin antisymmetric tensor defined as $\sigma_{\mu\nu} = 
i(\gamma_{\mu}\gamma_{\nu} - \gamma_{\nu}\gamma_{\mu}) / 2$, and
$\sigma \cdot T$ denotes $\sigma_{\mu\nu}T^{\mu\nu}$. Differentiating
Eq.~(\ref{eq:ECA}) with respect to the external source 
$T$, we can evaluate the VEV as follows: 
\begin{equation}
\label{eq:VEV0}
\langle i\psi^{\dagger}\sigma_{\mu\nu}\psi\rangle_F
=\frac{1}{VN_f}\frac{\partial\ln\mathcal{Z}_{\mathrm{eff}}[T,\hat{\mu}]}
{\partial T_{\mu\nu}}\Bigg|_{T=0}
=i\int\frac{d^4p}{(2\pi)^4}\mathrm{Tr}_{c,\gamma}
[S(P,\hat{\mu})\,\sigma_{\mu\nu}],
\end{equation}
where $P$ stands for $p-e_qV$. Equation~(\ref{eq:VEV0}) can be
evaluated further straightforwardly by expanding with respect to the
electric charge to order $\mathcal{O}(e_q)$ as follows:
\begin{eqnarray}
\label{eq:Ssi}
S(P,\mu)\,\sigma_{\mu\nu}
= \frac{1}{\rlap{/}{\bar{P}}+iM(\bar{P})}\sigma_{\mu\nu}
= -\frac{\frac{ie_q}{2}\sigma\cdot F
+i[\rlap{/}{P},M(\bar{P})]}{[\bar{p}^2+M^2(\bar{p})]^2}
[\bar{p}-iM(\bar{p})]\sigma_{\mu\nu}+\mathcal{O}(e^2_q).
\end{eqnarray}
In evaluating Eq.~(\ref{eq:Ssi}), we have used
the simplified notation $\bar{p}=p+i\hat{\mu}$ and the identity
$[P_{\mu},P_{\nu}]=ie_qF_{\mu\nu}$. The commutation relation in
Eq.~(\ref{eq:Ssi}) reads:  
\begin{eqnarray}
\label{eq:xsxs}
[\rlap{/}{P},M(\bar{P})]&=&\gamma_{\nu}[P_{\nu},M(\bar{P})]=
-2ie_q\tilde{M}'(\bar{p})\,(p_\mu F_{\mu\nu}+i\mu
F_{4\nu})\gamma_{\nu}. 
\end{eqnarray}
Note that there appears a pure electric part proportional to
$F_{4\nu}$ in the above equation, due to the breakdown of Lorentz
invariance at finite $\mu$ in addition to the usual EM
field-strength tensor, $F_{\mu\nu}$.  The $\tilde{M}'$  designates the
derivative of the effective-quark mass with respect to the momentum
squared:  
\begin{equation}
\label{eq:aaaaaa}
\tilde{M}'(\bar{p})=\frac{\partial M(\bar{p})}{\partial p^2}. 
\end{equation}
For convenience, we call the terms proportional to $\tilde{M}'$ as
a nonlocal contribution, since they arise from the nonlocal (derivative)   
interaction, whereas the terms without it a local one. Considering all
the ingredients discussed so far, and performing the trace of
Eq.~(\ref{eq:VEV0}) over the colour and Dirac spin spaces, we obtain as 
follows:   
\begin{eqnarray}
\label{eq:trs}
\mathrm{Tr}_{c,\gamma}[S\sigma_{\mu\nu}]
&=&\frac{4N_ce_qM(\bar{p})}{[\bar{p}^2+M^2(\bar{p})]^2}F_{\mu\nu}
-\frac{8N_ce_q\tilde{M}'(\bar{p})}{[\bar{p}^2+M^2(\bar{p})]^2}
\left[\bar{p}_{\mu}\bar{p}_\sigma
  F_{\sigma\nu}-(\mu\leftrightarrow\nu)\right] 
\nonumber\\
&\approx&
\underbrace{\frac{4N_ce_qM(\bar{p})}{[\bar{p}^2+M^2(\bar{p})]^2}F_{\mu\nu}
-\frac{8N_ce_q\tilde{M}'(\bar{p})}{[\bar{p}^2+M^2(\bar{p})]^2}
\left[p_{\mu}p_\sigma F_{\sigma\nu}-p_{\nu}p_\sigma
  F_{\sigma\mu}\right]}_\mathrm{magnetic+electric}
\nonumber\\
&-&\underbrace{\frac{8N_ce_q\tilde{M}'(\bar{p})}{[\bar{p}^2+M^2(\bar{p})]^2}
\left[\mu^2(\delta_{\nu4} F_{4\mu}-\delta_{\mu4}
  F_{4\nu})\right]}_\mathrm{electric}.
\end{eqnarray}
Here, we have ignored the terms proportional to $\mathcal{O}(p)$ in
evaluation of Eq.~(\ref{eq:trs}), since they become negligible
according to the integral identity $\int d^4p\,p_{\mu}f(p^2)=0$.  

First, we turn off the pure electric part in Eq.~(\ref{eq:trs}) for
simplicity, resulting in the magnetic contribution to the magnetic
susceptibility $\chi$, assigned as $\chi_\mathrm{M}$. Incorporating
Eqs.~(\ref{eq:VEV0}) and (\ref{eq:trs}), and then equating with
Eq.~(\ref{eq:VEV}), we finally arrive at the following compact
relation for $\chi_\mathrm{M}$:  
\begin{equation}
\label{eq:VEV2}
\chi_\mathrm{M}{\langle{i q}^{\dagger}q\rangle}=
4N_c \int\frac{d^4p}{(2\pi)^4}
\left[\frac{M(\bar{p})-\bar{p}^2\tilde{M}'(\bar{p})} 
{[\bar{p}^2+M^2(\bar{p})]^2}\right].
\end{equation}
One can easily see that Eq.~(\ref{eq:VEV2}) is a generalized
expression for Eq.~(52) given in Ref.~\cite{Kim:2004hd} by 
replacing $p\to p+i\mu$. Switching on the pure electric part, we can
write the electric contribution to the magnetic susceptibility as
$\chi_\mathrm{E}$ with an additional nonlocal term, coming from the
breakdown of Lorentz invariance, as follows:  
\begin{equation}
\label{eq:VEV3}
\chi_\mathrm{E}{\langle{i\psi}^{\dagger}\psi\rangle}
=\chi_\mathrm{M}{\langle{i\psi}^{\dagger}\psi\rangle}
+\underbrace{4N_c\int\frac{d^4p}{(2\pi)^4}
\Bigg[\frac{\mu^2\tilde{M}'(\bar{p})}{[\bar{p}^2 +
  M^2(\bar{p})]^2}\Bigg]}_\mathrm{Breakdown\,of\,Lorentz\,invariance}.
\end{equation}

\begin{figure}[t]
\begin{tabular}{cc}
\includegraphics[width=7.5cm]{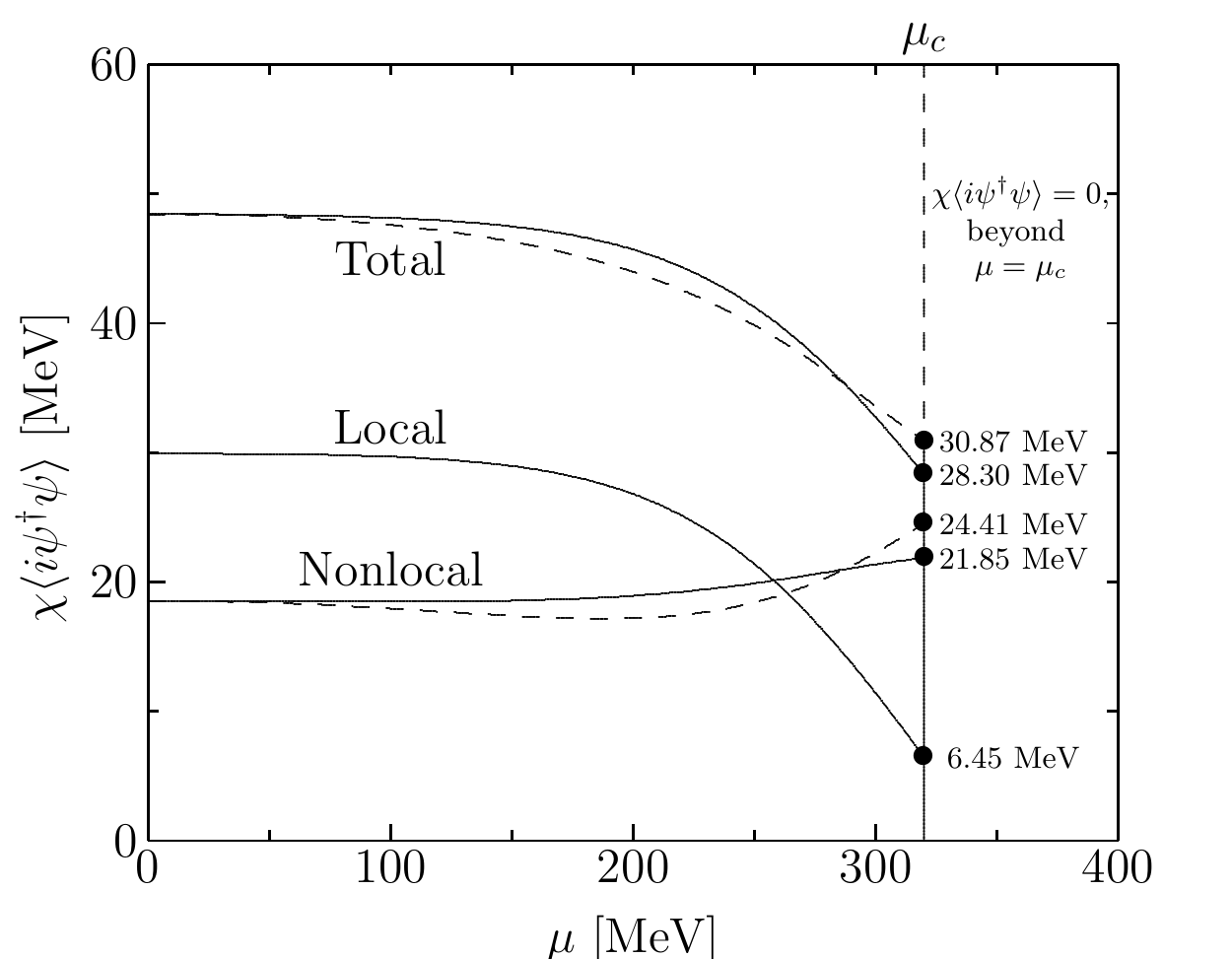}
\includegraphics[width=7.5cm]{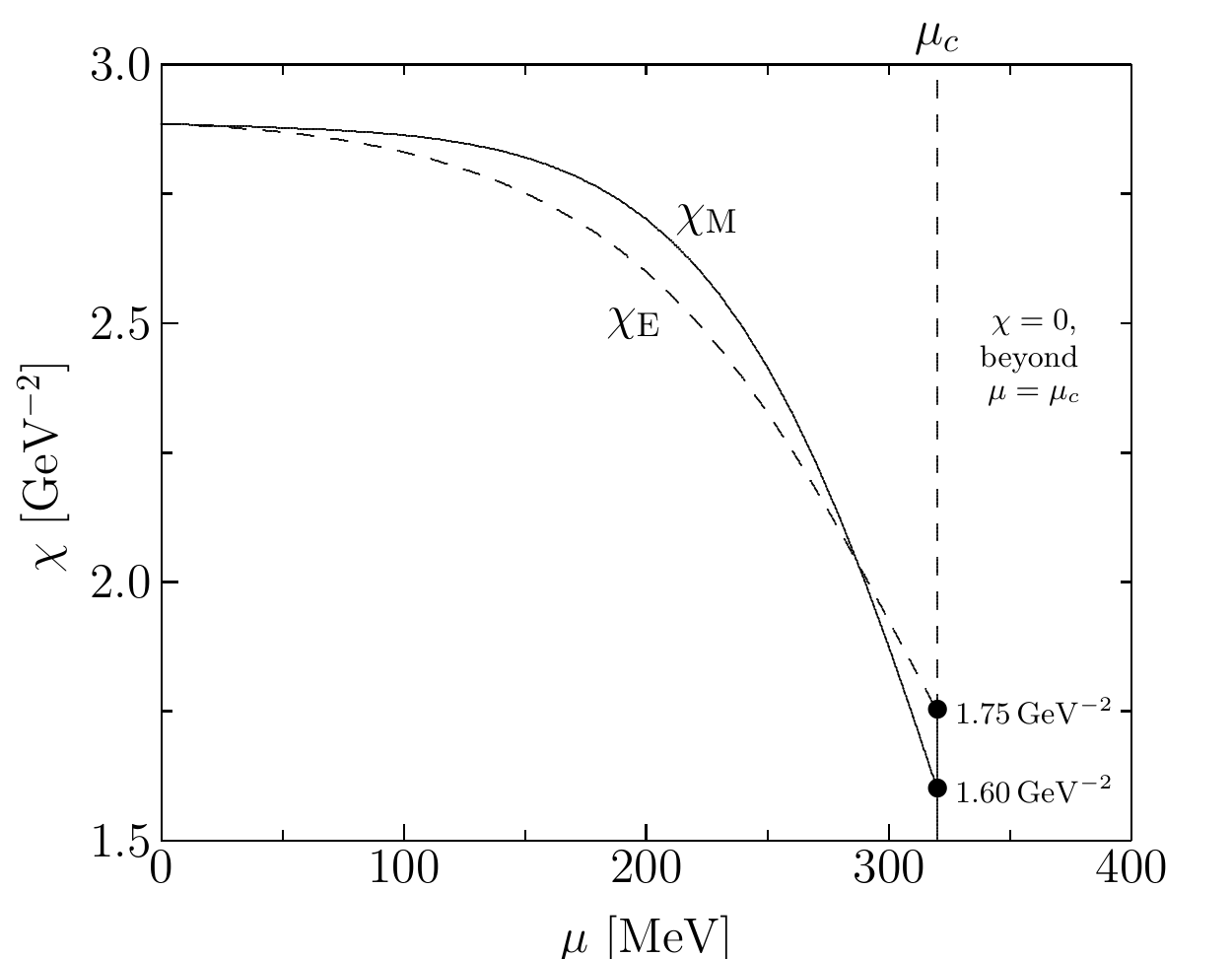}
\end{tabular}
\caption{$\chi_\mathrm{M,E}\langle i\psi^{\dagger}\psi\rangle$ (left
  panel) and $\chi_\mathrm{M,E}$ (right panel) as functions of
  $\mu$. The magnetic contribution is presented in the solid curve,
  whereas the electric one in the dashed one.  The vertical dashed
  lines indicate the $\mu_c\approx320$ MeV.}      
\label{fig1}
\end{figure}

In Fig.~\ref{fig1}, we present the numerical results for the
$\chi_\mathrm{M,E}\langle i\psi^{\dagger}\psi\rangle$ (left panel) and
the $\chi_\mathrm{M,E}$ (right panel) as functions of $\mu$,
respectively.  The magnetic contribution is drawn in the solid curve,
whereas the electric one in the dashed one.  As for the
$\chi_\mathrm{M,E}\langle i\psi^{\dagger}\psi\rangle$ depicted in the
left panel, we show each contribution (local, nonlocal and total)
separately. Note that the dominant local contributions are the same
for the magnetic and electric contributions as expected from
Eqs.~(\ref{eq:VEV2}) and (\ref{eq:VEV3}), and starts to decrease
beyond $\mu\approx200$ MeV. At $\mu=\mu_c$, its strength is reduced to 
about a half of that at $\mu=0$.  On the contrary, the nonlocal
contributions are almost flat for the NG phase, and show {\it small}
difference between the magnetic and electric ones, caused by the
breakdown of Lorentz invariance, being different from the pion
weak decay constant~\cite{Nam:2008xx,Kim:2003tp,Kirchbach:1997rk}. As
a result, the total contributions resemble the local one.  Moreover,
the {\it first-order} magnetic phase transition occurs at the critical
chemical potential, following the chiral restoration, as shown in 
Fig.~\ref{fig1}.  Several model calculations, using the nonlocal
chiral quark model~\cite{Kim:2004hd} and QCD sum
rules~\cite{Belyaev:1984ic,Balitsky:1985aq,Ball:2002ps}, computed  
$\chi\langle i\psi^{\dagger}\psi\rangle$ at $\mu=0$. We note that our
value $\sim46$ MeV is in good agreement with those in
Refs.~~\cite{Belyaev:1984ic,Balitsky:1985aq,Ball:2002ps}. 

Finally, in the right panel of Fig.~\ref{fig1}, we show the numerical
results for the $\chi_\mathrm{M,E}$ in the same manner as in the left 
panel. Since the chiral condensate is rather flat for the NG phase
with respect to $\mu$ as shown in
Refs.~\cite{Carter:1998ji,Nam:2008xx}, the overall tendency of the
$\chi_\mathrm{M,E}$ is very similar to that of the
$\chi_\mathrm{M,E}\langle i\psi^{\dagger}\psi\rangle$.  Again, we
observe small difference between the $\chi_\mathrm{M}$ and
$\chi_\mathrm{E}$, and the first-order magnetic phase transition at
the critical point.  Note that in the vicinity of $\mu_c$ the 
$\chi_\mathrm{M,E}$ drecreases by about a factor of two 
 in magnitude compared to its value at $\mu=0$.  As mentioned
previously, since the Fermi surface begins to be filled just beyond
$\mu\approx300$ MeV, which corresponds approximately to the normal
nuclear density $\rho_0\approx0.17\,\mathrm{fm}^{-3}$, we 
expect that there is a rapid change of the $\chi$ near $\rho_0$.  In
other words, the response of the QCD vacuum becomes weak but unstable
to the external EM source as the density is closed to $\rho_0$, while
it remains relatively stable otherwise.  

In Ref.~\cite{Tatsumi:2008nx}, the colour-symmetric one-gluon exchange
(OGE) was employed to investigate the $\chi$ in medium in addition to
the static screening effect, and was found that a sharp divergence
takes place in the $\chi$ near the normal nuclear density, although it 
depends on the screening and the current-quark mass. This observation
may be consistent with the present results showing the first-order
magnetic phase transition.  

In fact, partial results of the magnetic susceptibility at finite quark-chemical potential has been already studies within the same
framework~\cite{Nam:2008}.  However, there are several important
points that were not considered in Ref.~\cite{Nam:2008}.  Firstly, the
breaking of Lorentz invariance was assumed to be small, so that it was
ignored.  Secondly, the chemical potential was treated perturbatively,
so that the calculation was performed by expanding it for numerical
simplicity.  Unfortunately, this naive expansion has led to the wrong
behavior of the magnetic susceptibility, in particular, as $\mu$ is
getting close to its critical value.  Finally, the finite value of
the current quark mass was taken into account in Ref.~\cite{Nam:2008}.
However, we soon realized  that a mere extension of the model by
Carter and Diakonov~\cite{Carter:1998ji} seems to be unjustified.
Because of this reason, we are restricted ourselves to consider the
chiral limit in the present work.

\section{Summary and conclusions}
In the present work, we have investigated the QCD magnetic
susceptibility $\chi$ induced by the external constant 
electromagnetic field at the finite quark-chemical 
potential, i.e. $\mu\ne0$ at $T=0$. We focussed on the Nambu-Goldstone
(NG) phase, in which the $\chi$ is well defined.  We employed the
$\mu$-modified  nonlocal chiral quark model derived from the instanton
vacuum at the normalization scale $\Lambda=\bar{\rho}^{-1}\approx600$
MeV, corresponding to the phenomenological instanton parameters
$\bar{R}\approx1$ fm and $\bar{\rho}\approx1/3$ fm.  

We started with the $\mu$-modified Dirac equation to obtain the quark
propagator in the instanton ensemble. The $\mu$-dependent effective
quark mass can be obtained from the Fourier transform of the
quark-zero mode solution. However, we employed the parameterized
dipole-type form factor instead of it to circumvent numerical
difficulties. The effective partition function was constructed in such
way to reproduce the quark propagator in the presence of the instanton
background.  Solving the Dyson-Schwinger-Gorkov equation, as done in 
Ref.~\cite{Carter:1998ji}, we were able to explore the phase structure
through two order parameters, $M_0$ and $\Delta$, corresponding 
to the Nambu-Goldstone and colour-superconducting phases,
respectively.  As a result, the critical quark-chemical potential $\mu_c$ was
determined to be about $320$ MeV. 

Using the low-energy effective partition function with the finite 
quark-chemical potential, we computed the magnetic susceptibility of
the QCD vacuum $\chi$.  We found that the $\chi$ remains stable up to 
$\mu\approx200$ MeV, and drops then drastically.  At the critical 
quark-chemical potential, $\mu_c\approx320$ MeV, the strength of the
$\chi$ is decreased by about a factor of two, compared to that at
$\mu=0$, and the first-order magnetic phase transition takes place,
corresponding to the chiral restoration.  From these observations, we
conclude that the response of the QCD vacuum becomes weak
(insensitive) and unstable to the external electromagnetic source near
the normal nuclear density in comparison to that for the vacuum.
Moreover, the effect of the breakdown of Lorentz invariance, caused by
the finite $\mu$, on the $\chi$ turns out to be small,
$\chi_\mathrm{M}\sim\chi_\mathrm{E}$: We note that this tendency is
rather different from the pion weak decay constant. Related works
including explicit flavour SU(3) symmetry 
breaking effects as well as meson-loop corrections at finite quark-chemical potential   
are under progress.   

We want at this point to mention that we have calculated the
magnetic susceptibility in the NG phase only, that is, below the
critical quark-chemical potential.  Note that there is, however, one
caveat: In principle, we can compute the VEV in Eq.~(\ref{eq:VEV})
beyond the critical quark-chemical potential, i.e., in the CSC phase.
We have to keep in mind that the SU(3) colour 
symmetry is broken down to $\mathrm{SU(2)}\times\mathrm{U(1)}$ in the
CSC pahse, which causes a lift of colour degeneracy.  Thus, the quark
propagator must be separated into two different quark propagators of
which each consists of $4\times 4$ propagators in the chiral 
$L$-$R$ basis~\cite{Carter:1998ji}.  The form of the propagators have
been obtained by solving the Schwinger-Dyson-Gorkov equation with a
systematic expansion in parameters $1/N_c$ and $\bar{\rho}/\bar{R}$.  
Taking consideration of these propagators, we can compute the magnetic
susceptibility beyond the NG phase, though the calculation is quite
involved.   

Last but not least, the magnetic susceptibility defined in
Eq.(\ref{eq:VEV}) is not the magnetization in medium as done in
Ref.~\cite{Tatsumi:2008nx}.  The magnetization can be rewritten as the
VEV of the following operator $\langle iq^\dagger \gamma_4
\sigma_{\mu\nu}q\rangle_F$.  This is another interesting quantity
which we can compute within the present scheme.  The related
investigations are under way. 
\section*{Acknowledgments}
Authors are grateful to T.~Tatsumi for invaluable discussions and
critical comments.  The authors thank T.~Kunihiro, S.~H.~Lee and C.~W.~Kao for
fruitful discussions.  The work of S.i.N. is partially supported by
the grant for Scientific Research (Priority Area No.17070002 and
No.20028005) from the Ministry of Education, Culture, Science and
Technology (MEXT) of Japan and the grant of NSC\,96-2112-M033-003-MY3 from the National Science Council (NSC) of Taiwan. The work of H.Ch.K. is supported by Inha University Research Grant (INHA-37453).  This work was also done
under the Yukawa International Program for Quark-Hadron Sciences. The
numerical calculations were carried out on YISUN at YITP in Kyoto
University.     

\end{document}